\documentclass[
reprint,
 amsmath,amssymb,
 aps,
]{revtex4-1}

\usepackage{graphicx}
\usepackage{dcolumn}
\usepackage{bm}
\usepackage{siunitx}

\begin{document}


\title{Optically Coupled Methods for Microwave Impedance Microscopy}

\author{Scott R. Johnston}
\author{Eric Yue Ma}
\author{Zhi-Xun Shen}
\affiliation{Department of Applied Physics, Stanford University, Stanford, Ca 94305, USA}%


\date{\today}
\begin{abstract}

Scanning Microwave Impedance Microscopy (MIM) measurement of photoconductivity with 50 nm resolution is demonstrated using a modulated optical source. The use of a modulated source allows for measurement of photoconductivity in a single scan without a reference region on the sample, as well as removing most topographical artifacts and enhancing signal to noise as compared with unmodulated measurement. A broadband light source with tunable monochrometer is then used to measure energy resolved photoconductivity with the same methodology. Finally, a pulsed optical source is used to measure local photo-carrier lifetimes via MIM, using the same 50 nm resolution tip.

\end{abstract}

\maketitle
%

\section{Introduction}

Scanning Microwave Impedance Microscopy (MIM) is a scanning probe microscopy technique which measures inhomogeneity in sample conductivity and permittivity capacitively, requiring no direct electrical connection to the sample \cite{Lai2007, Lai2008b}. The technique can achieve a spatial resolution of 50 nm \cite{Nanotubes2016}, determined predominately by tip size.

MIM has been applied successfully to measure conductivity inhomogeneity in a wide variety of systems, including carbon nano-tubes \cite{Nanotubes2016}, graphene \cite{Kundhikanjana2009}, In$_2$Se$_3$ nanoribbons \cite{Lai2009a}, quantum hall edge states \cite{Lai2011}, MoS$_2$ field effect transistors \cite{Wu2016}, and more \cite{Lai2010a, Tselev2012, Kundhikanjana2013, Tselev2014, Ma2015b, Ma2015c, Ma2015d, Kundhikanjana2015, Berweger2016, Tselev2016, Biagi2016}.

MIM has also been used to measure photovoltaics and other light sensitive materials, including lead halide perovskites \cite{Chu, Berweger2017}, CdTe \cite{Tuteja2015}, and monolayer WS$_2$/WS$_{2(1 - x)}$Se$_{2x}$ heterostructures \cite{Tsai2017}. For most of these studies \cite{Chu, Tuteja2015, Tsai2017} a comparison is made between dark and illuminated MIM scans to study the spatially resolved photoconductivity of the material. This approach suffers from many drawbacks, however; the sample must be scanned multiple times, and any change in tip condition or other measurement conditions between the scans can create erroneous differences. By measuring many times at different optical powers and demonstrating a clear trend \cite{Chu, Tsai2017} these difficulties may be largely overcome.

Other optically coupled scanning probe techniques include photoconductive atomic force microscopy (AFM) \cite{Coffey2007, Dang2010} and  photovoltaic Kelvin probe force microscopy (KPFM) \cite{Tennyson2015, Tennyson2016, Garrett2017}. Photoconductive AFM has the same drawbacks as standard conductive AFM, as compared with MIM. These include the need for an ohmic contact with the sample, and high sensitivity to the tip-sample interface contact. Photovoltaic KPFM does not involve tip-sample contact, but measures photo-voltage, rather than photoconductivity, and requires grounding the back of the sample relative to which the top surface potential is measured.

We propose and demonstrate a different approach to optically coupled MIM measurement, in which the optical power is rapidly varied ($\geqslant$ 1kHz) and the MIM signal is demodulated at this optical modulation frequency. This approach is related to other modulated scanning capacitance measurement schemes such as scanning nonlinear dielectric microscopy \cite{Cho2011} and scanning capacitance microscopy \cite{Williams1989}. Both of those measurements use a low frequency (kHz - MHz) electric field modulation and measure the varying capacitance at a higher frequency (GHz). This results in a measurement of $dC/dV$, the change in capacitance under applied field. Another modulated scanning capacitance method, tuning fork MIM, modulates the distance between the tip and the sample using a quartz tuning fork \cite{Cui2016}, measuring change in capacitance as a function of tip height, $dC/dz$, much like electrostatic force microscopy (EFM), but with the additional information provided by a GHz measurement of both capacitance and loss.

Modulating the optical source allows us to obtain spatially resolved measurements of photoconductivity, $d\sigma/dL$, where $L$ is the light intensity at the tip-sample interface, in a single scan without electrical contacts. The noise rejection of the lock-in measurement also permits photoconductivity measurements done with a weak, variable-wavelength source such as a tungsten lamp with a monochrometer, which would otherwise produce a signal too weak for standard MIM. 

Finally, we implemented time-resolved optically-coupled MIM, in which an optical pulse is repeatedly applied to the sample and the subsequent MIM response is averaged over repetitions. This allows spatially and time resolved measurement of photoconductivity, which can be used to determine carrier lifetimes. The measurement of carrier lifetimes is comparable to time-resolved photoluminescence, but at higher spatial resolution and lower temporal resolution.

\section{Optically Modulated MIM}

\begin{figure*}
\includegraphics[width = 17cm]{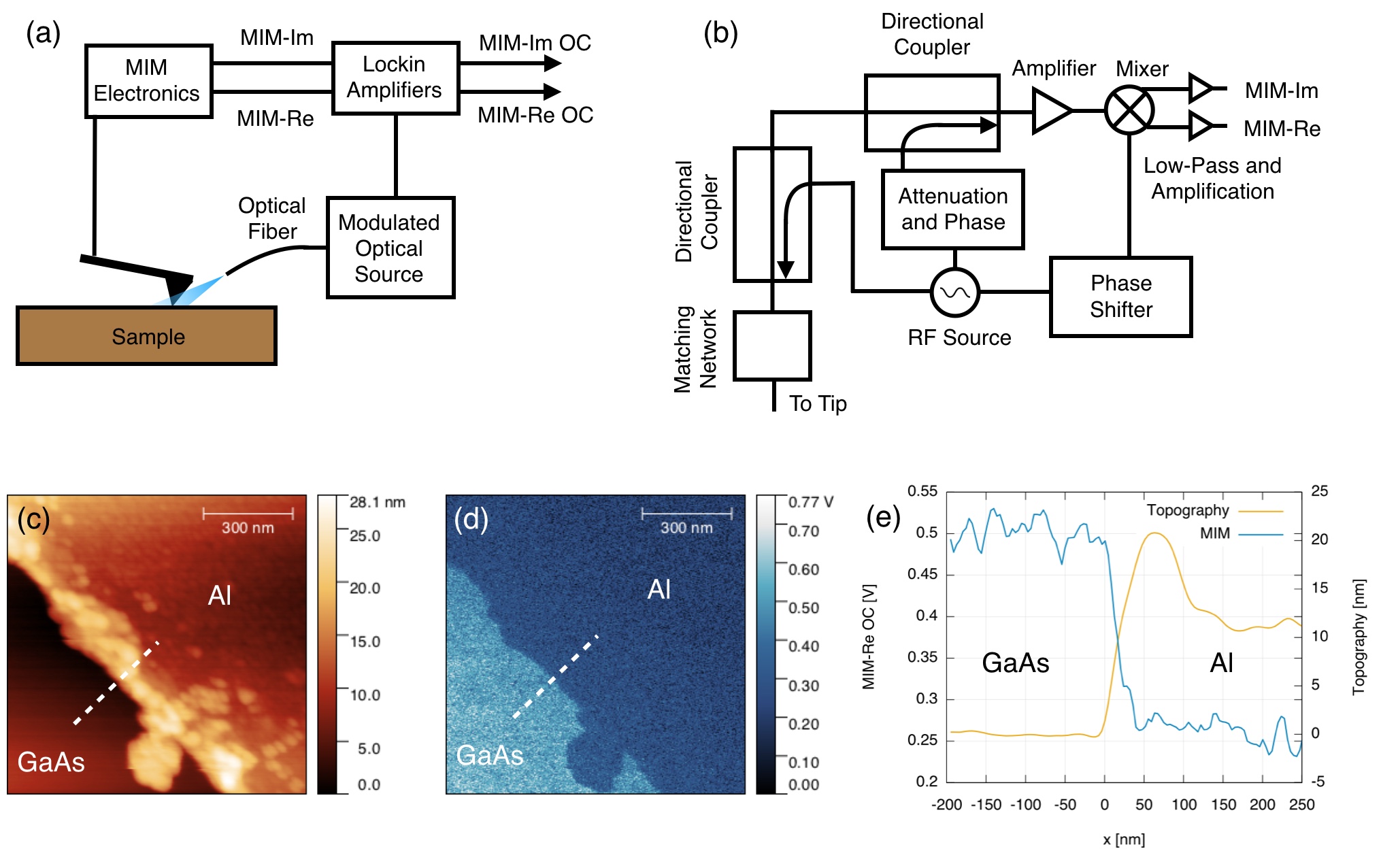}
\caption{
Optically modulated MIM. (a) Measurement Block Diagram. (b) Schematic of MIM Electronics (c) Topography of test sample ($\approx$ 12 nm Al on GaAs). (d) Optically modulated MIM-Re signal. (e) Line-cut of optically modulated MIM-Re signal showing $\approx$ 50 nm spatial resolution.
}
\label{fig1}
\end{figure*}

Optically modulated MIM measurement is done by applying a modulated optical source to the sample, then demodulating the MIM signal at the optical modulation frequency, as shown in Fig. \ref{fig1} a. For these measurements, a fiber-coupled 455 nm LED was used, with modulation supplied by a high-speed LED controller driven with a square wave at 3 kHz. The bare fiber end is brought within a few mm of the tip-sample interface at a roughly 30$^\circ$ angle to the sample surface, illuminating a few mm$^2$ around the tip-sample contact.

The working principle of the MIM measurement is that a $\approx$ 1 GHz microwave signal is reflected from the tip, and variations in the tip-sample impedance change the amplitude and phase of the reflected signal, allowing measurement of changes in sample conductivity and permittivity. These changes in conductivity can be a function of position, or a different parameter. In this paper, we use MIM to measure the change in conductivity as the level of optical illumination is varied. Fig. \ref{fig1} b gives an overview of the electronics used to measure these small changes in reflected microwave signal. 

The LED current is set such that the power output from the fiber is 1 mW as measured by a thermopile optical power sensor. We found that this power is sufficiently low to produce minimal thermal expansion effects, related to those seen in photoconductive AFM \cite{sviridov2016}. Unlike in photoconductive AFM, MIM is not sensitive to the contact force. The thermally induced modulation of cantilever-sample distance can create a significant background at high enough optical power levels, however, beginning at $\approx 3$ mW. 
One might also wonder whether the top illumination would interfere with the laser used for topographical feedback, however this laser operates in the infra-red, far away from the spectrum used for optical excitation of the sample. Thus, we found no significant effect of the top illumination on the laser topographical feedback system.

The optical modulation results in a signal which corresponds to the change in local conductivity under applied light, or local photoconductivity (Fig. \ref{fig1} d). The optically coupled MIM signal (MIM-OC) is much smaller on the Al surface where there is no photoconductivity, with the residual signal (background) resulting from coupling of the cantilever and upper tip to the GaAs regions of the sample. In the GaAs region the super-band-gap light creates carriers, increasing the conductivity, which leads to the photoconductivity measured by MIM-OC. The MIM-OC signal is strikingly uniform in both the Al and GaAs regions, with no drift or topography coupling, unlike what is typically seen in MIM measurements. 

The largest source of topographical artifacts in unmodulated MIM is typically due to variation in cantilever-sample capacitance as the tip moves vertically to follow sample topography. The variation in cantilever-sample distance creates topographical artifacts in the measured admittance which are proportional to $z$, the sample height at the tip. 


With optically modulated MIM, this change in absolute capacitance is no longer measured; only the modulated capacitance is measured. While there is some optically modulated signal picked up by the cantilever, resulting in a small background signal, this signal is not changed significantly by topography, as observed in Fig. \ref{fig1} d. Local topography can still affect optically modulated MIM measurements, however, both through local geometry (eg. $\nabla^2 z$) \cite{Berweger2017} and inhomogeneous lighting. To demonstrate this, optically modulated MIM measurements of a highly textured, photoconductive sample are shown in Appendix \ref{texture}. 

The transition between Al and GaAs is remarkably sharp, showing $<$50 nm resolution (Fig. \ref{fig1} e). This can be attributed to the use of a 20 nm Rocky Mountain Nanotechnology etched platinum probe \cite{Bussmann2004}.

\section{Energy Resolved Optical MIM}

\begin{figure}
\includegraphics[width = 8.5cm]{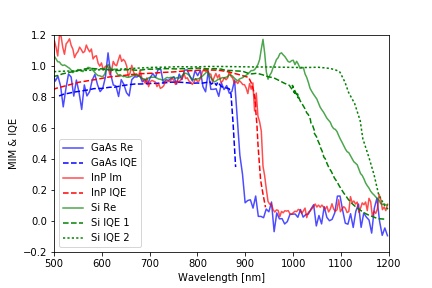}
\caption{
Energy resolved MIM for various semidconductors and their corresponding internal quantum efficiencies (IQEs) from literature:  GaAs \cite{Xiong2010}, InP \cite{111606}, Si 1 \cite{Liang2015}, Si 2 \cite{Zhao1996}. Multiple sources are cited for Si IQE because of differences in IR IQE due to device structure. Each measurement is taken at a single point on the sample, with MIM-OC measured while the optical wavelength is swept. All MIM signals have been normalized over wavelength to optical intensity (Appendix \ref{intensity}), and have been scaled to match the absolute values of the IQE.
}
\label{fig2}
\end{figure}

By varying the wavelength of the modulated light source it is possible to measure the photoconductivity of the material as a function of not only position but also photon energy. For this experiment a tungsten lamp and a monochrometer with a chopper at its entrance slit are used as the modulated light-source. Spatial scans can be done at any wavelength using this setup, though the scan time must be significantly longer than when the LED source is used due to the $\mu$W power level. 

Using the tunable wavelength source we can do single point spectroscopy, as shown in Fig. \ref{fig2}. Here the tip is held at a single spot on the sample and the wavelength is varied. All energy-resolved data must be normalized by the power spectrum of the source, as shown in Appendix \ref{intensity}. 

For all samples, the MIM channel (real or imaginary) with the greatest signal level is shown. Typical MIM response as a function of conductivity is shown in \cite{Kundhikanjana2009}. Our measurements depend on the derivative of that MIM response with respect to conductivity. Thus, at low conductivities ($>10^3$ $\Omega$cm) the real channel dominates, white at more intermediate conductivities ($\approx 10^2$ $\Omega$cm) we expect the imaginary channel to be stronger than the real channel.

In a simple semiconducting system, such as those we studied here, the change in conductivity at fixed intensity over wavelength will be proportion to the internal quantum efficiency (IQE). This is because change in conductivity is given by: $$\Delta \sigma = e \Delta p (\mu_n + \mu_p) $$ where $\Delta p$ is the change in electron-hole pair concentration. The recombination rate, $r_{\text{recombination}}$ is proportional to the pair density, $p$, and the pair generation rate, $r_{\text{generation}}$, is proportional to the quantum efficiency, $QE_\lambda$. Since these are equal in equilibrium, we get:
 $$p \propto r_{\text{recombination}} = r_{\text{generation}} \propto QE_\lambda$$ 
 Altogether: 
 $$\Delta \sigma \propto QE_\lambda$$ 
 Thus, the energy-resolved optical MIM signal is closely related to the IQE, as shown in Fig. \ref{fig2}. 

There is some variation in the literature measurements of IQE for the IR range of Si, where the intermediate IQE values can be enhanced by device structure \cite{Holman2013}. Hence, we show only that our measured IQE for Si falls within the typical range, whereas the measurements for our direct band-gap materials, GaAs and InP, show far less variation.

This technique provides a way to spatially resolve any variation in band-gap energy with higher spatial resolution than is typically achievable by purely optical methods such as photoluminescence or reflectivity, which are limited by diffraction.

\section{Time Resolved Optical MIM}

\begin{figure}
\includegraphics[width = 8cm]{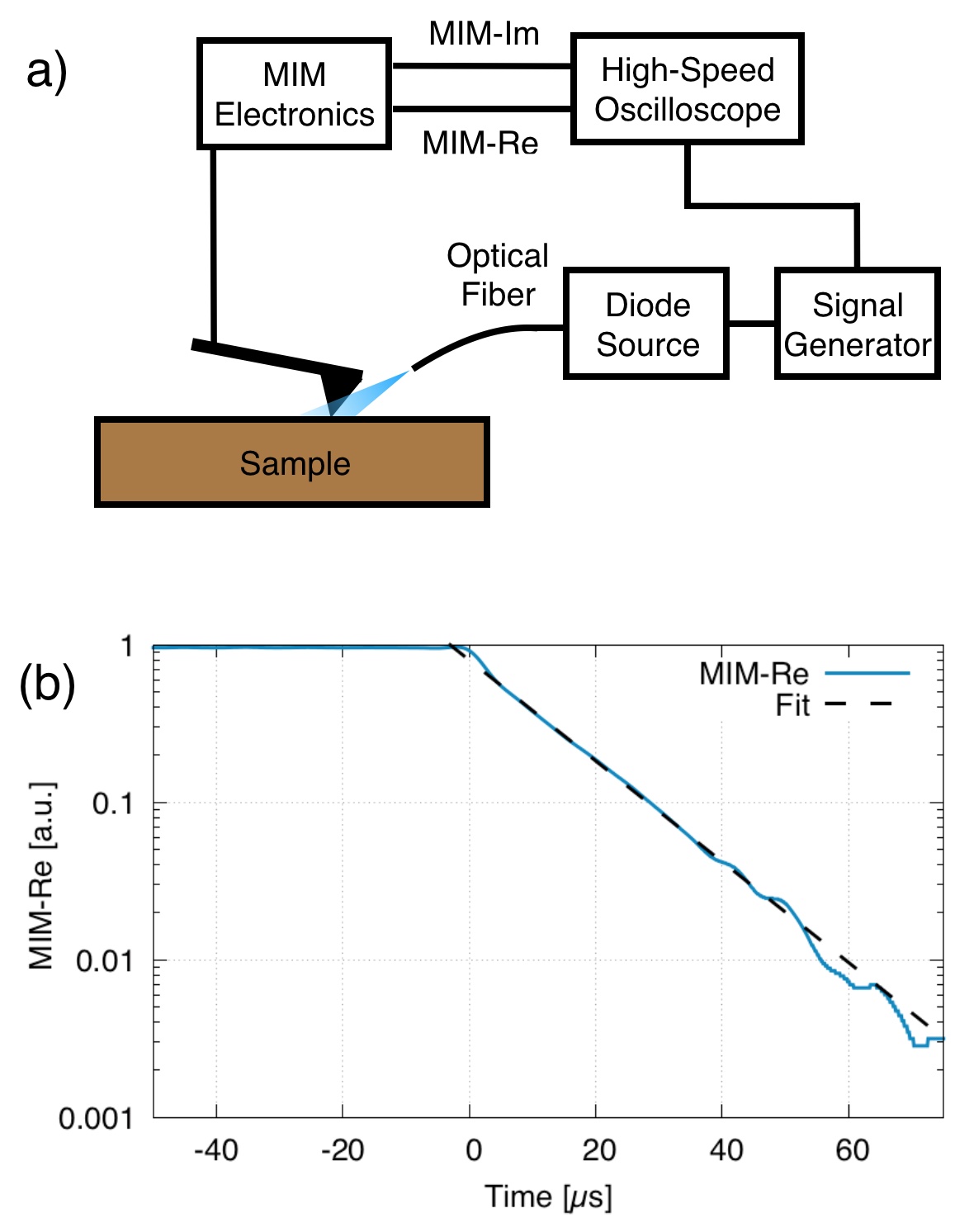}
\caption{
Time resolved MIM. (a) Measurement setup. (b) Time resolved MIM signal from undoped Si. Fitting gives a time constant $\tau$ = 13.6 $\mu$s, consistent with literature values \cite{Kiliani2011}.
}
\label{fig3}
\end{figure}

In this section, we use MIM to time resolve the conductivity change immediately after a pulse of light is applied or removed. This allows us to measure the carrier lifetime in the material with the same 50 nm spatial resolution as our previous measurements. This type of measurement is also commonly done by photoluminescence \cite{Kiliani2011}, which has relatively limited spatial resolution, though better time resolution.

The optical pulse train used for this measurement is the same as that for optically modulated MIM measurement, a 3 kHz square wave; however, instead of using a lockin amplifier, the resulting MIM signal is measured with a high-speed oscilloscope synced to the pulse generator (Fig. \ref{fig3} a). The oscilloscope is used to average together thousands of time-traces immediately after the falling edge of the applied optical pulse, providing a time-resolved measurement of the decrease in conductivity after sample illumination is removed.

For these measurements, we used a fiber-coupled LED light source with a high speed (1ns) diode driver. We measured this light source to provide a 20ns rise time with 40ns fall time. The 20ns filter on the output of the MIM electronics also limits the bandwidth of the measurement. However, for photovoltaic materials with carrier lifetimes of 100s of ns or microseconds, the time resolution of the instrument and light source are not a limiting factor.

We observed different signal levels and time-constants from different materials, with the high signal level and long time-constant of undoped Si providing very clear exponential behavior over multiple orders of magnitude (Fig. \ref{fig3} b). Fitting to the Si time response gives a time constant of 13.6 $\mu$s, which falls into the center of the typical distribution of lifetimes seen in undoped Si \cite{Kiliani2011}.\\

\section{Discussion}

We have demonstrated not only the ability to use MIM to clearly and consistently measure absolute photoconductivity with 50 nm spatial resolution, but also to measure IQE as a function of photon energy, and to measure local carrier lifetime. Combined, these methods promise to be a powerful tool to probe local effects in photo-sensitive materials, such as band-gap and carrier life-time variation near grain boundaries in perovskites \cite{Chu, Berweger2017} and other polycrystalline materials \cite{Tuteja2015}, as well as edge effects in 2D semiconductors \cite{Wu2016}.

%

\section*{Acknowledgement}

The authors would like to acknowledge the support of NSF grant DMR1305731 and The Gordon and Betty Moore Foundation, EPiOs grant GBMF4536. We also thank Patrick Kirchmann and Abraham Saldivar for their technical support. Z.-X.S. is a cofounder of PrimeNano Inc., which licensed MIM technology from Stanford for commercial instrumentation.

\bibliography{library}

\begin{thebibliography}{38}%
\makeatletter
\providecommand \@ifxundefined [1]{%
 \@ifx{#1\undefined}
}%
\providecommand \@ifnum [1]{%
 \ifnum #1\expandafter \@firstoftwo
 \else \expandafter \@secondoftwo
 \fi
}%
\providecommand \@ifx [1]{%
 \ifx #1\expandafter \@firstoftwo
 \else \expandafter \@secondoftwo
 \fi
}%
\providecommand \natexlab [1]{#1}%
\providecommand \enquote  [1]{``#1''}%
\providecommand \bibnamefont  [1]{#1}%
\providecommand \bibfnamefont [1]{#1}%
\providecommand \citenamefont [1]{#1}%
\providecommand \href@noop [0]{\@secondoftwo}%
\providecommand \href [0]{\begingroup \@sanitize@url \@href}%
\providecommand \@href[1]{\@@startlink{#1}\@@href}%
\providecommand \@@href[1]{\endgroup#1\@@endlink}%
\providecommand \@sanitize@url [0]{\catcode `\\12\catcode `\$12\catcode
  `\&12\catcode `\#12\catcode `\^12\catcode `\_12\catcode `\%12\relax}%
\providecommand \@@startlink[1]{}%
\providecommand \@@endlink[0]{}%
\providecommand \url  [0]{\begingroup\@sanitize@url \@url }%
\providecommand \@url [1]{\endgroup\@href {#1}{\urlprefix }}%
\providecommand \urlprefix  [0]{URL }%
\providecommand \Eprint [0]{\href }%
\providecommand \doibase [0]{http://dx.doi.org/}%
\providecommand \selectlanguage [0]{\@gobble}%
\providecommand \bibinfo  [0]{\@secondoftwo}%
\providecommand \bibfield  [0]{\@secondoftwo}%
\providecommand \translation [1]{[#1]}%
\providecommand \BibitemOpen [0]{}%
\providecommand \bibitemStop [0]{}%
\providecommand \bibitemNoStop [0]{.\EOS\space}%
\providecommand \EOS [0]{\spacefactor3000\relax}%
\providecommand \BibitemShut  [1]{\csname bibitem#1\endcsname}%
\let\auto@bib@innerbib\@empty
\bibitem [{\citenamefont {Lai}\ \emph {et~al.}(2007)\citenamefont {Lai},
  \citenamefont {Ji}, \citenamefont {Leindecker}, \citenamefont {Kelly},\ and\
  \citenamefont {Shen}}]{Lai2007}%
  \BibitemOpen
  \bibfield  {author} {\bibinfo {author} {\bibfnamefont {K.}~\bibnamefont
  {Lai}}, \bibinfo {author} {\bibfnamefont {M.~B.}\ \bibnamefont {Ji}},
  \bibinfo {author} {\bibfnamefont {N.}~\bibnamefont {Leindecker}}, \bibinfo
  {author} {\bibfnamefont {M.~A.}\ \bibnamefont {Kelly}}, \ and\ \bibinfo
  {author} {\bibfnamefont {Z.~X.}\ \bibnamefont {Shen}},\ }\href {\doibase
  10.1063/1.2746768} {\bibfield  {journal} {\bibinfo  {journal} {Rev. Sci.
  Instrum.}\ }\textbf {\bibinfo {volume} {78}} (\bibinfo {year} {2007}),\
  10.1063/1.2746768},\ \Eprint {http://arxiv.org/abs/0703382} {arXiv:0703382
  [cond-mat]} \BibitemShut {NoStop}%
\bibitem [{\citenamefont {Lai}\ \emph {et~al.}(2008)\citenamefont {Lai},
  \citenamefont {Kundhikanjana}, \citenamefont {Kelly},\ and\ \citenamefont
  {Shen}}]{Lai2008b}%
  \BibitemOpen
  \bibfield  {author} {\bibinfo {author} {\bibfnamefont {K.}~\bibnamefont
  {Lai}}, \bibinfo {author} {\bibfnamefont {W.}~\bibnamefont {Kundhikanjana}},
  \bibinfo {author} {\bibfnamefont {M.}~\bibnamefont {Kelly}}, \ and\ \bibinfo
  {author} {\bibfnamefont {Z.~X.}\ \bibnamefont {Shen}},\ }\href {\doibase
  10.1063/1.2949109} {\bibfield  {journal} {\bibinfo  {journal} {Rev. Sci.
  Instrum.}\ }\textbf {\bibinfo {volume} {79}} (\bibinfo {year} {2008}),\
  10.1063/1.2949109}\BibitemShut {NoStop}%
\bibitem [{\citenamefont {Seabron}\ \emph {et~al.}(2016)\citenamefont
  {Seabron}, \citenamefont {Maclaren}, \citenamefont {Xie}, \citenamefont
  {Rotkin}, \citenamefont {Rogers},\ and\ \citenamefont
  {Wilson}}]{Nanotubes2016}%
  \BibitemOpen
  \bibfield  {author} {\bibinfo {author} {\bibfnamefont {E.}~\bibnamefont
  {Seabron}}, \bibinfo {author} {\bibfnamefont {S.}~\bibnamefont {Maclaren}},
  \bibinfo {author} {\bibfnamefont {X.}~\bibnamefont {Xie}}, \bibinfo {author}
  {\bibfnamefont {S.~V.}\ \bibnamefont {Rotkin}}, \bibinfo {author}
  {\bibfnamefont {J.~A.}\ \bibnamefont {Rogers}}, \ and\ \bibinfo {author}
  {\bibfnamefont {W.~L.}\ \bibnamefont {Wilson}},\ }\href {\doibase
  10.1021/acsnano.5b04975} {\bibfield  {journal} {\bibinfo  {journal} {ACS
  Nano}\ }\textbf {\bibinfo {volume} {10}},\ \bibinfo {pages} {360} (\bibinfo
  {year} {2016})}\BibitemShut {NoStop}%
\bibitem [{\citenamefont {Kundhikanjana}\ \emph {et~al.}(2009)\citenamefont
  {Kundhikanjana}, \citenamefont {Lai}, \citenamefont {Wang}, \citenamefont
  {Dai}, \citenamefont {Kelly},\ and\ \citenamefont
  {Shen}}]{Kundhikanjana2009}%
  \BibitemOpen
  \bibfield  {author} {\bibinfo {author} {\bibfnamefont {W.}~\bibnamefont
  {Kundhikanjana}}, \bibinfo {author} {\bibfnamefont {K.}~\bibnamefont {Lai}},
  \bibinfo {author} {\bibfnamefont {H.}~\bibnamefont {Wang}}, \bibinfo {author}
  {\bibfnamefont {H.}~\bibnamefont {Dai}}, \bibinfo {author} {\bibfnamefont
  {M.~A.}\ \bibnamefont {Kelly}}, \ and\ \bibinfo {author} {\bibfnamefont
  {Z.-X.}\ \bibnamefont {Shen}},\ }\href {\doibase 10.1021/nl901949z}
  {\bibfield  {journal} {\bibinfo  {journal} {Nano Lett.}\ }\textbf {\bibinfo
  {volume} {9}},\ \bibinfo {pages} {3762} (\bibinfo {year} {2009})}\BibitemShut
  {NoStop}%
\bibitem [{\citenamefont {Lai}\ \emph {et~al.}(2009)\citenamefont {Lai},
  \citenamefont {Peng}, \citenamefont {Kundhikanjana}, \citenamefont {Schoen},
  \citenamefont {Xie}, \citenamefont {Meister}, \citenamefont {Cui},
  \citenamefont {Kelly},\ and\ \citenamefont {Shen}}]{Lai2009a}%
  \BibitemOpen
  \bibfield  {author} {\bibinfo {author} {\bibfnamefont {K.}~\bibnamefont
  {Lai}}, \bibinfo {author} {\bibfnamefont {H.}~\bibnamefont {Peng}}, \bibinfo
  {author} {\bibfnamefont {W.}~\bibnamefont {Kundhikanjana}}, \bibinfo {author}
  {\bibfnamefont {D.~T.}\ \bibnamefont {Schoen}}, \bibinfo {author}
  {\bibfnamefont {C.}~\bibnamefont {Xie}}, \bibinfo {author} {\bibfnamefont
  {S.}~\bibnamefont {Meister}}, \bibinfo {author} {\bibfnamefont
  {Y.}~\bibnamefont {Cui}}, \bibinfo {author} {\bibfnamefont {M.~A.}\
  \bibnamefont {Kelly}}, \ and\ \bibinfo {author} {\bibfnamefont {Z.~X.}\
  \bibnamefont {Shen}},\ }\href {\doibase 10.1021/nl900222j} {\bibfield
  {journal} {\bibinfo  {journal} {Nano Lett.}\ }\textbf {\bibinfo {volume}
  {9}},\ \bibinfo {pages} {1265} (\bibinfo {year} {2009})},\ \Eprint
  {http://arxiv.org/abs/0902.2255} {arXiv:0902.2255} \BibitemShut {NoStop}%
\bibitem [{\citenamefont {Lai}\ \emph {et~al.}(2011)\citenamefont {Lai},
  \citenamefont {Kundhikanjana}, \citenamefont {Kelly}, \citenamefont {Shen},
  \citenamefont {Shabani},\ and\ \citenamefont {Shayegan}}]{Lai2011}%
  \BibitemOpen
  \bibfield  {author} {\bibinfo {author} {\bibfnamefont {K.}~\bibnamefont
  {Lai}}, \bibinfo {author} {\bibfnamefont {W.}~\bibnamefont {Kundhikanjana}},
  \bibinfo {author} {\bibfnamefont {M.~A.}\ \bibnamefont {Kelly}}, \bibinfo
  {author} {\bibfnamefont {Z.~X.}\ \bibnamefont {Shen}}, \bibinfo {author}
  {\bibfnamefont {J.}~\bibnamefont {Shabani}}, \ and\ \bibinfo {author}
  {\bibfnamefont {M.}~\bibnamefont {Shayegan}},\ }\href
  {http://link.aps.org/doi/10.1103/PhysRevLett.107.176809} {\bibfield
  {journal} {\bibinfo  {journal} {Phys. Rev. Lett.}\ }\textbf {\bibinfo
  {volume} {107}},\ \bibinfo {pages} {176809} (\bibinfo {year}
  {2011})}\BibitemShut {NoStop}%
\bibitem [{\citenamefont {Wu}\ \emph {et~al.}(2016)\citenamefont {Wu},
  \citenamefont {Li}, \citenamefont {Luan}, \citenamefont {Wu}, \citenamefont
  {Li}, \citenamefont {Yogeesh}, \citenamefont {Ghosh}, \citenamefont {Chu},
  \citenamefont {Akinwande}, \citenamefont {Niu},\ and\ \citenamefont
  {Lai}}]{Wu2016}%
  \BibitemOpen
  \bibfield  {author} {\bibinfo {author} {\bibfnamefont {D.}~\bibnamefont
  {Wu}}, \bibinfo {author} {\bibfnamefont {X.}~\bibnamefont {Li}}, \bibinfo
  {author} {\bibfnamefont {L.}~\bibnamefont {Luan}}, \bibinfo {author}
  {\bibfnamefont {X.}~\bibnamefont {Wu}}, \bibinfo {author} {\bibfnamefont
  {W.}~\bibnamefont {Li}}, \bibinfo {author} {\bibfnamefont {M.~N.}\
  \bibnamefont {Yogeesh}}, \bibinfo {author} {\bibfnamefont {R.}~\bibnamefont
  {Ghosh}}, \bibinfo {author} {\bibfnamefont {Z.}~\bibnamefont {Chu}}, \bibinfo
  {author} {\bibfnamefont {D.}~\bibnamefont {Akinwande}}, \bibinfo {author}
  {\bibfnamefont {Q.}~\bibnamefont {Niu}}, \ and\ \bibinfo {author}
  {\bibfnamefont {K.}~\bibnamefont {Lai}},\ }\href {\doibase
  10.1073/pnas.1605982113} {\bibfield  {journal} {\bibinfo  {journal} {PNAS}\
  }\textbf {\bibinfo {volume} {113}},\ \bibinfo {pages} {8583} (\bibinfo {year}
  {2016})}\BibitemShut {NoStop}%
\bibitem [{\citenamefont {Lai}\ \emph {et~al.}(2010)\citenamefont {Lai},
  \citenamefont {Nakamura}, \citenamefont {Kundhikanjana}, \citenamefont
  {Kawasaki}, \citenamefont {Tokura}, \citenamefont {Kelly},\ and\
  \citenamefont {Shen}}]{Lai2010a}%
  \BibitemOpen
  \bibfield  {author} {\bibinfo {author} {\bibfnamefont {K.}~\bibnamefont
  {Lai}}, \bibinfo {author} {\bibfnamefont {M.}~\bibnamefont {Nakamura}},
  \bibinfo {author} {\bibfnamefont {W.}~\bibnamefont {Kundhikanjana}}, \bibinfo
  {author} {\bibfnamefont {M.}~\bibnamefont {Kawasaki}}, \bibinfo {author}
  {\bibfnamefont {Y.}~\bibnamefont {Tokura}}, \bibinfo {author} {\bibfnamefont
  {M.~A.}\ \bibnamefont {Kelly}}, \ and\ \bibinfo {author} {\bibfnamefont
  {Z.-X.}\ \bibnamefont {Shen}},\ }\href {\doibase 10.1126/science.1189925}
  {\bibfield  {journal} {\bibinfo  {journal} {Science}\ }\textbf {\bibinfo
  {volume} {329}},\ \bibinfo {pages} {190} (\bibinfo {year} {2010})},\ \Eprint
  {http://arxiv.org/abs/0006190} {arXiv:0006190 [cond-mat]} \BibitemShut
  {NoStop}%
\bibitem [{\citenamefont {Tselev}\ \emph {et~al.}(2012)\citenamefont {Tselev},
  \citenamefont {Lavrik}, \citenamefont {Vlassiouk}, \citenamefont {Briggs},
  \citenamefont {Rutgers}, \citenamefont {Proksch},\ and\ \citenamefont
  {Kalinin}}]{Tselev2012}%
  \BibitemOpen
  \bibfield  {author} {\bibinfo {author} {\bibfnamefont {A.}~\bibnamefont
  {Tselev}}, \bibinfo {author} {\bibfnamefont {N.~V.}\ \bibnamefont {Lavrik}},
  \bibinfo {author} {\bibfnamefont {I.}~\bibnamefont {Vlassiouk}}, \bibinfo
  {author} {\bibfnamefont {D.~P.}\ \bibnamefont {Briggs}}, \bibinfo {author}
  {\bibfnamefont {M.}~\bibnamefont {Rutgers}}, \bibinfo {author} {\bibfnamefont
  {R.}~\bibnamefont {Proksch}}, \ and\ \bibinfo {author} {\bibfnamefont
  {S.~V.}\ \bibnamefont {Kalinin}},\ }\href {\doibase
  10.1088/0957-4484/23/38/385706} {\bibfield  {journal} {\bibinfo  {journal}
  {Nanotechnology}\ }\textbf {\bibinfo {volume} {23}} (\bibinfo {year}
  {2012}),\ 10.1088/0957-4484/23/38/385706}\BibitemShut {NoStop}%
\bibitem [{\citenamefont {Kundhikanjana}\ \emph {et~al.}(2013)\citenamefont
  {Kundhikanjana}, \citenamefont {Yang}, \citenamefont {Tanga}, \citenamefont
  {Zhang}, \citenamefont {Lai}, \citenamefont {Ma}, \citenamefont {Kelly},
  \citenamefont {Li},\ and\ \citenamefont {Shen}}]{Kundhikanjana2013}%
  \BibitemOpen
  \bibfield  {author} {\bibinfo {author} {\bibfnamefont {W.}~\bibnamefont
  {Kundhikanjana}}, \bibinfo {author} {\bibfnamefont {Y.}~\bibnamefont {Yang}},
  \bibinfo {author} {\bibfnamefont {Q.}~\bibnamefont {Tanga}}, \bibinfo
  {author} {\bibfnamefont {K.}~\bibnamefont {Zhang}}, \bibinfo {author}
  {\bibfnamefont {K.}~\bibnamefont {Lai}}, \bibinfo {author} {\bibfnamefont
  {Y.}~\bibnamefont {Ma}}, \bibinfo {author} {\bibfnamefont {M.~A.}\
  \bibnamefont {Kelly}}, \bibinfo {author} {\bibfnamefont {X.~X.}\ \bibnamefont
  {Li}}, \ and\ \bibinfo {author} {\bibfnamefont {Z.-X.}\ \bibnamefont
  {Shen}},\ }\href {\doibase 10.1088/0268-1242/28/2/025010} {\bibfield
  {journal} {\bibinfo  {journal} {Semicond. Sci. Technol.}\ }\textbf {\bibinfo
  {volume} {28}},\ \bibinfo {pages} {025010} (\bibinfo {year}
  {2013})}\BibitemShut {NoStop}%
\bibitem [{\citenamefont {Tselev}\ \emph {et~al.}(2014)\citenamefont {Tselev},
  \citenamefont {Ivanov}, \citenamefont {Lavrik}, \citenamefont {Belianinov},
  \citenamefont {Jesse}, \citenamefont {Mathews}, \citenamefont {Mitchell},\
  and\ \citenamefont {Kalinin}}]{Tselev2014}%
  \BibitemOpen
  \bibfield  {author} {\bibinfo {author} {\bibfnamefont {A.}~\bibnamefont
  {Tselev}}, \bibinfo {author} {\bibfnamefont {I.~N.}\ \bibnamefont {Ivanov}},
  \bibinfo {author} {\bibfnamefont {N.~V.}\ \bibnamefont {Lavrik}}, \bibinfo
  {author} {\bibfnamefont {A.}~\bibnamefont {Belianinov}}, \bibinfo {author}
  {\bibfnamefont {S.}~\bibnamefont {Jesse}}, \bibinfo {author} {\bibfnamefont
  {J.~P.}\ \bibnamefont {Mathews}}, \bibinfo {author} {\bibfnamefont {G.~D.}\
  \bibnamefont {Mitchell}}, \ and\ \bibinfo {author} {\bibfnamefont {S.~V.}\
  \bibnamefont {Kalinin}},\ }\href {\doibase 10.1016/j.fuel.2014.02.029}
  {\bibfield  {journal} {\bibinfo  {journal} {Fuel}\ }\textbf {\bibinfo
  {volume} {126}},\ \bibinfo {pages} {32} (\bibinfo {year} {2014})}\BibitemShut
  {NoStop}%
\bibitem [{\citenamefont {Ma}\ \emph {et~al.}(2015{\natexlab{a}})\citenamefont
  {Ma}, \citenamefont {Calvo}, \citenamefont {Wang}, \citenamefont {Lian},
  \citenamefont {M{\"{u}}hlbauer}, \citenamefont {Br{\"{u}}ne}, \citenamefont
  {Cui}, \citenamefont {Lai}, \citenamefont {Kundhikanjana}, \citenamefont
  {Yang}, \citenamefont {Baenninger}, \citenamefont {K{\"{o}}nig},
  \citenamefont {Ames}, \citenamefont {Buhmann}, \citenamefont {Leubner},
  \citenamefont {Molenkamp}, \citenamefont {Zhang}, \citenamefont
  {Goldhaber-Gordon}, \citenamefont {Kelly},\ and\ \citenamefont
  {Shen}}]{Ma2015b}%
  \BibitemOpen
  \bibfield  {author} {\bibinfo {author} {\bibfnamefont {E.~Y.}\ \bibnamefont
  {Ma}}, \bibinfo {author} {\bibfnamefont {M.~R.}\ \bibnamefont {Calvo}},
  \bibinfo {author} {\bibfnamefont {J.}~\bibnamefont {Wang}}, \bibinfo {author}
  {\bibfnamefont {B.}~\bibnamefont {Lian}}, \bibinfo {author} {\bibfnamefont
  {M.}~\bibnamefont {M{\"{u}}hlbauer}}, \bibinfo {author} {\bibfnamefont
  {C.}~\bibnamefont {Br{\"{u}}ne}}, \bibinfo {author} {\bibfnamefont {Y.-T.}\
  \bibnamefont {Cui}}, \bibinfo {author} {\bibfnamefont {K.}~\bibnamefont
  {Lai}}, \bibinfo {author} {\bibfnamefont {W.}~\bibnamefont {Kundhikanjana}},
  \bibinfo {author} {\bibfnamefont {Y.}~\bibnamefont {Yang}}, \bibinfo {author}
  {\bibfnamefont {M.}~\bibnamefont {Baenninger}}, \bibinfo {author}
  {\bibfnamefont {M.}~\bibnamefont {K{\"{o}}nig}}, \bibinfo {author}
  {\bibfnamefont {C.}~\bibnamefont {Ames}}, \bibinfo {author} {\bibfnamefont
  {H.}~\bibnamefont {Buhmann}}, \bibinfo {author} {\bibfnamefont
  {P.}~\bibnamefont {Leubner}}, \bibinfo {author} {\bibfnamefont {L.~W.}\
  \bibnamefont {Molenkamp}}, \bibinfo {author} {\bibfnamefont {S.-C.}\
  \bibnamefont {Zhang}}, \bibinfo {author} {\bibfnamefont {D.}~\bibnamefont
  {Goldhaber-Gordon}}, \bibinfo {author} {\bibfnamefont {M.~A.}\ \bibnamefont
  {Kelly}}, \ and\ \bibinfo {author} {\bibfnamefont {Z.-X.}\ \bibnamefont
  {Shen}},\ }\href {\doibase 10.1038/ncomms8252} {\bibfield  {journal}
  {\bibinfo  {journal} {Nat. Commun.}\ }\textbf {\bibinfo {volume} {6}},\
  \bibinfo {pages} {7252} (\bibinfo {year} {2015}{\natexlab{a}})},\ \Eprint
  {http://arxiv.org/abs/1212.6441} {arXiv:1212.6441} \BibitemShut {NoStop}%
\bibitem [{\citenamefont {Ma}\ \emph {et~al.}(2015{\natexlab{b}})\citenamefont
  {Ma}, \citenamefont {Bryant}, \citenamefont {Tokunaga}, \citenamefont
  {Aeppli}, \citenamefont {Tokura},\ and\ \citenamefont {Shen}}]{Ma2015c}%
  \BibitemOpen
  \bibfield  {author} {\bibinfo {author} {\bibfnamefont {E.~Y.}\ \bibnamefont
  {Ma}}, \bibinfo {author} {\bibfnamefont {B.}~\bibnamefont {Bryant}}, \bibinfo
  {author} {\bibfnamefont {Y.}~\bibnamefont {Tokunaga}}, \bibinfo {author}
  {\bibfnamefont {G.}~\bibnamefont {Aeppli}}, \bibinfo {author} {\bibfnamefont
  {Y.}~\bibnamefont {Tokura}}, \ and\ \bibinfo {author} {\bibfnamefont {Z.-X.}\
  \bibnamefont {Shen}},\ }\href {\doibase 10.1038/ncomms8595} {\bibfield
  {journal} {\bibinfo  {journal} {Nat. Commun.}\ }\textbf {\bibinfo {volume}
  {6}},\ \bibinfo {pages} {7595} (\bibinfo {year}
  {2015}{\natexlab{b}})}\BibitemShut {NoStop}%
\bibitem [{\citenamefont {Ma}\ \emph {et~al.}(2015{\natexlab{c}})\citenamefont
  {Ma}, \citenamefont {Cui}, \citenamefont {Ueda}, \citenamefont {Tang},
  \citenamefont {Chen}, \citenamefont {Tamura}, \citenamefont {Wu},
  \citenamefont {Fujioka}, \citenamefont {Tokura},\ and\ \citenamefont
  {Shen}}]{Ma2015d}%
  \BibitemOpen
  \bibfield  {author} {\bibinfo {author} {\bibfnamefont {E.~Y.}\ \bibnamefont
  {Ma}}, \bibinfo {author} {\bibfnamefont {Y.-T.}\ \bibnamefont {Cui}},
  \bibinfo {author} {\bibfnamefont {K.}~\bibnamefont {Ueda}}, \bibinfo {author}
  {\bibfnamefont {S.}~\bibnamefont {Tang}}, \bibinfo {author} {\bibfnamefont
  {K.}~\bibnamefont {Chen}}, \bibinfo {author} {\bibfnamefont {N.}~\bibnamefont
  {Tamura}}, \bibinfo {author} {\bibfnamefont {P.~M.}\ \bibnamefont {Wu}},
  \bibinfo {author} {\bibfnamefont {J.}~\bibnamefont {Fujioka}}, \bibinfo
  {author} {\bibfnamefont {Y.}~\bibnamefont {Tokura}}, \ and\ \bibinfo {author}
  {\bibfnamefont {Z.-X.}\ \bibnamefont {Shen}},\ }\href {\doibase
  10.1126/science.aac8289} {\bibfield  {journal} {\bibinfo  {journal}
  {Science}\ }\textbf {\bibinfo {volume} {350}},\ \bibinfo {pages} {538}
  (\bibinfo {year} {2015}{\natexlab{c}})},\ \Eprint
  {http://arxiv.org/abs/1011.1669v3} {arXiv:1011.1669v3} \BibitemShut {NoStop}%
\bibitem [{\citenamefont {Kundhikanjana}\ \emph {et~al.}(2015)\citenamefont
  {Kundhikanjana}, \citenamefont {Sheng}, \citenamefont {Yang}, \citenamefont
  {Lai}, \citenamefont {Ma}, \citenamefont {Cui}, \citenamefont {Kelly},
  \citenamefont {Nakamura}, \citenamefont {Kawasaki}, \citenamefont {Tokura},
  \citenamefont {Tang}, \citenamefont {Zhang}, \citenamefont {Li},\ and\
  \citenamefont {Shen}}]{Kundhikanjana2015}%
  \BibitemOpen
  \bibfield  {author} {\bibinfo {author} {\bibfnamefont {W.}~\bibnamefont
  {Kundhikanjana}}, \bibinfo {author} {\bibfnamefont {Z.}~\bibnamefont
  {Sheng}}, \bibinfo {author} {\bibfnamefont {Y.}~\bibnamefont {Yang}},
  \bibinfo {author} {\bibfnamefont {K.}~\bibnamefont {Lai}}, \bibinfo {author}
  {\bibfnamefont {E.~Y.}\ \bibnamefont {Ma}}, \bibinfo {author} {\bibfnamefont
  {Y.~T.}\ \bibnamefont {Cui}}, \bibinfo {author} {\bibfnamefont {M.~A.}\
  \bibnamefont {Kelly}}, \bibinfo {author} {\bibfnamefont {M.}~\bibnamefont
  {Nakamura}}, \bibinfo {author} {\bibfnamefont {M.}~\bibnamefont {Kawasaki}},
  \bibinfo {author} {\bibfnamefont {Y.}~\bibnamefont {Tokura}}, \bibinfo
  {author} {\bibfnamefont {Q.}~\bibnamefont {Tang}}, \bibinfo {author}
  {\bibfnamefont {K.}~\bibnamefont {Zhang}}, \bibinfo {author} {\bibfnamefont
  {X.}~\bibnamefont {Li}}, \ and\ \bibinfo {author} {\bibfnamefont {Z.~X.}\
  \bibnamefont {Shen}},\ }\href {\doibase 10.1103/PhysRevLett.115.265701}
  {\bibfield  {journal} {\bibinfo  {journal} {Phys. Rev. Lett.}\ }\textbf
  {\bibinfo {volume} {115}},\ \bibinfo {pages} {1} (\bibinfo {year} {2015})},\
  \Eprint {http://arxiv.org/abs/1306.3065} {arXiv:1306.3065} \BibitemShut
  {NoStop}%
\bibitem [{\citenamefont {Berweger}\ \emph {et~al.}(2016)\citenamefont
  {Berweger}, \citenamefont {Blanchard}, \citenamefont {Quardokus},
  \citenamefont {Delrio}, \citenamefont {Wallis}, \citenamefont {Kabos},
  \citenamefont {Krylyuk},\ and\ \citenamefont {Davydov}}]{Berweger2016}%
  \BibitemOpen
  \bibfield  {author} {\bibinfo {author} {\bibfnamefont {S.}~\bibnamefont
  {Berweger}}, \bibinfo {author} {\bibfnamefont {P.~T.}\ \bibnamefont
  {Blanchard}}, \bibinfo {author} {\bibfnamefont {R.~C.}\ \bibnamefont
  {Quardokus}}, \bibinfo {author} {\bibfnamefont {F.~W.}\ \bibnamefont
  {Delrio}}, \bibinfo {author} {\bibfnamefont {T.~M.}\ \bibnamefont {Wallis}},
  \bibinfo {author} {\bibfnamefont {P.}~\bibnamefont {Kabos}}, \bibinfo
  {author} {\bibfnamefont {S.}~\bibnamefont {Krylyuk}}, \ and\ \bibinfo
  {author} {\bibfnamefont {A.~V.}\ \bibnamefont {Davydov}},\ }in\ \href
  {\doibase 10.1109/MWSYM.2016.7540184} {\emph {\bibinfo {booktitle} {IEEE
  MTT-S Int. Microw. Symp. Dig.}}}\ (\bibinfo {year} {2016})\BibitemShut
  {NoStop}%
\bibitem [{\citenamefont {Tselev}\ \emph {et~al.}(2016)\citenamefont {Tselev},
  \citenamefont {Yu}, \citenamefont {Cao}, \citenamefont {Dedon}, \citenamefont
  {Martin}, \citenamefont {Kalinin},\ and\ \citenamefont
  {Maksymovych}}]{Tselev2016}%
  \BibitemOpen
  \bibfield  {author} {\bibinfo {author} {\bibfnamefont {A.}~\bibnamefont
  {Tselev}}, \bibinfo {author} {\bibfnamefont {P.}~\bibnamefont {Yu}}, \bibinfo
  {author} {\bibfnamefont {Y.}~\bibnamefont {Cao}}, \bibinfo {author}
  {\bibfnamefont {L.~R.}\ \bibnamefont {Dedon}}, \bibinfo {author}
  {\bibfnamefont {L.~W.}\ \bibnamefont {Martin}}, \bibinfo {author}
  {\bibfnamefont {S.~V.}\ \bibnamefont {Kalinin}}, \ and\ \bibinfo {author}
  {\bibfnamefont {P.}~\bibnamefont {Maksymovych}},\ }\href {\doibase
  10.1038/ncomms11630} {\bibfield  {journal} {\bibinfo  {journal} {Nat.
  Commun.}\ }\textbf {\bibinfo {volume} {7}},\ \bibinfo {pages} {11630}
  (\bibinfo {year} {2016})}\BibitemShut {NoStop}%
\bibitem [{\citenamefont {Biagi}\ \emph {et~al.}(2016)\citenamefont {Biagi},
  \citenamefont {Fabregas}, \citenamefont {Gramse}, \citenamefont {{Van Der
  Hofstadt}}, \citenamefont {Ju{\'{a}}rez}, \citenamefont {Kienberger},
  \citenamefont {Fumagalli},\ and\ \citenamefont {Gomila}}]{Biagi2016}%
  \BibitemOpen
  \bibfield  {author} {\bibinfo {author} {\bibfnamefont {M.~C.}\ \bibnamefont
  {Biagi}}, \bibinfo {author} {\bibfnamefont {R.}~\bibnamefont {Fabregas}},
  \bibinfo {author} {\bibfnamefont {G.}~\bibnamefont {Gramse}}, \bibinfo
  {author} {\bibfnamefont {M.}~\bibnamefont {{Van Der Hofstadt}}}, \bibinfo
  {author} {\bibfnamefont {A.}~\bibnamefont {Ju{\'{a}}rez}}, \bibinfo {author}
  {\bibfnamefont {F.}~\bibnamefont {Kienberger}}, \bibinfo {author}
  {\bibfnamefont {L.}~\bibnamefont {Fumagalli}}, \ and\ \bibinfo {author}
  {\bibfnamefont {G.}~\bibnamefont {Gomila}},\ }\href {\doibase
  10.1021/acsnano.5b04279} {\bibfield  {journal} {\bibinfo  {journal} {ACS
  Nano}\ }\textbf {\bibinfo {volume} {10}},\ \bibinfo {pages} {280} (\bibinfo
  {year} {2016})}\BibitemShut {NoStop}%
\bibitem [{\citenamefont {Chu}\ \emph {et~al.}(2016)\citenamefont {Chu},
  \citenamefont {Yang}, \citenamefont {Schulz}, \citenamefont {Wu},
  \citenamefont {Ma}, \citenamefont {Seifert}, \citenamefont {Sun},
  \citenamefont {Zhu}, \citenamefont {Li},\ and\ \citenamefont {Lai}}]{Chu}%
  \BibitemOpen
  \bibfield  {author} {\bibinfo {author} {\bibfnamefont {Z.}~\bibnamefont
  {Chu}}, \bibinfo {author} {\bibfnamefont {M.}~\bibnamefont {Yang}}, \bibinfo
  {author} {\bibfnamefont {P.}~\bibnamefont {Schulz}}, \bibinfo {author}
  {\bibfnamefont {D.}~\bibnamefont {Wu}}, \bibinfo {author} {\bibfnamefont
  {X.}~\bibnamefont {Ma}}, \bibinfo {author} {\bibfnamefont {E.}~\bibnamefont
  {Seifert}}, \bibinfo {author} {\bibfnamefont {L.}~\bibnamefont {Sun}},
  \bibinfo {author} {\bibfnamefont {K.}~\bibnamefont {Zhu}}, \bibinfo {author}
  {\bibfnamefont {X.}~\bibnamefont {Li}}, \ and\ \bibinfo {author}
  {\bibfnamefont {K.}~\bibnamefont {Lai}},\ }\href@noop {} {\bibfield
  {journal} {\bibinfo  {journal} {arXiv:1610.00755}\ } (\bibinfo {year}
  {2016})}\BibitemShut {NoStop}%
\bibitem [{\citenamefont {Berweger}\ \emph {et~al.}(2017)\citenamefont
  {Berweger}, \citenamefont {MacDonald}, \citenamefont {Yang}, \citenamefont
  {Coakley}, \citenamefont {Berry}, \citenamefont {Zhu}, \citenamefont
  {DelRio}, \citenamefont {Wallis},\ and\ \citenamefont
  {Kabos}}]{Berweger2017}%
  \BibitemOpen
  \bibfield  {author} {\bibinfo {author} {\bibfnamefont {S.}~\bibnamefont
  {Berweger}}, \bibinfo {author} {\bibfnamefont {G.~A.}\ \bibnamefont
  {MacDonald}}, \bibinfo {author} {\bibfnamefont {M.}~\bibnamefont {Yang}},
  \bibinfo {author} {\bibfnamefont {K.~J.}\ \bibnamefont {Coakley}}, \bibinfo
  {author} {\bibfnamefont {J.~J.}\ \bibnamefont {Berry}}, \bibinfo {author}
  {\bibfnamefont {K.}~\bibnamefont {Zhu}}, \bibinfo {author} {\bibfnamefont
  {F.~W.}\ \bibnamefont {DelRio}}, \bibinfo {author} {\bibfnamefont {T.~M.}\
  \bibnamefont {Wallis}}, \ and\ \bibinfo {author} {\bibfnamefont
  {P.}~\bibnamefont {Kabos}},\ }\href {\doibase 10.1021/acs.nanolett.6b05119}
  {\bibfield  {journal} {\bibinfo  {journal} {Nano Lett.}\ }\textbf {\bibinfo
  {volume} {17}},\ \bibinfo {pages} {1796} (\bibinfo {year}
  {2017})}\BibitemShut {NoStop}%
\bibitem [{\citenamefont {Tuteja}\ \emph {et~al.}(2015)\citenamefont {Tuteja},
  \citenamefont {Koirala}, \citenamefont {Maclaren}, \citenamefont {Collins},\
  and\ \citenamefont {Rockett}}]{Tuteja2015}%
  \BibitemOpen
  \bibfield  {author} {\bibinfo {author} {\bibfnamefont {M.}~\bibnamefont
  {Tuteja}}, \bibinfo {author} {\bibfnamefont {P.}~\bibnamefont {Koirala}},
  \bibinfo {author} {\bibfnamefont {S.}~\bibnamefont {Maclaren}}, \bibinfo
  {author} {\bibfnamefont {R.}~\bibnamefont {Collins}}, \ and\ \bibinfo
  {author} {\bibfnamefont {A.}~\bibnamefont {Rockett}},\ }\href {\doibase
  10.1063/1.4932952} {\bibfield  {journal} {\bibinfo  {journal} {Appl. Phys.
  Lett.}\ }\textbf {\bibinfo {volume} {107}} (\bibinfo {year} {2015}),\
  10.1063/1.4932952}\BibitemShut {NoStop}%
\bibitem [{\citenamefont {Tsai}\ \emph {et~al.}(2017)\citenamefont {Tsai},
  \citenamefont {Chu}, \citenamefont {Han}, \citenamefont {Chuu}, \citenamefont
  {Wu}, \citenamefont {Johnson}, \citenamefont {Cheng}, \citenamefont {Chou},
  \citenamefont {Muller}, \citenamefont {Li}, \citenamefont {Lai},\ and\
  \citenamefont {Shih}}]{Tsai2017}%
  \BibitemOpen
  \bibfield  {author} {\bibinfo {author} {\bibfnamefont {Y.}~\bibnamefont
  {Tsai}}, \bibinfo {author} {\bibfnamefont {Z.}~\bibnamefont {Chu}}, \bibinfo
  {author} {\bibfnamefont {Y.}~\bibnamefont {Han}}, \bibinfo {author}
  {\bibfnamefont {C.-P.}\ \bibnamefont {Chuu}}, \bibinfo {author}
  {\bibfnamefont {D.}~\bibnamefont {Wu}}, \bibinfo {author} {\bibfnamefont
  {A.}~\bibnamefont {Johnson}}, \bibinfo {author} {\bibfnamefont
  {F.}~\bibnamefont {Cheng}}, \bibinfo {author} {\bibfnamefont {M.-Y.}\
  \bibnamefont {Chou}}, \bibinfo {author} {\bibfnamefont {D.~A.}\ \bibnamefont
  {Muller}}, \bibinfo {author} {\bibfnamefont {X.}~\bibnamefont {Li}}, \bibinfo
  {author} {\bibfnamefont {K.}~\bibnamefont {Lai}}, \ and\ \bibinfo {author}
  {\bibfnamefont {C.-K.}\ \bibnamefont {Shih}},\ }\href {\doibase
  10.1002/adma.201703680} {\bibfield  {journal} {\bibinfo  {journal} {Adv.
  Mater.}\ }\textbf {\bibinfo {volume} {29}},\ \bibinfo {pages} {1703680}
  (\bibinfo {year} {2017})}\BibitemShut {NoStop}%
\bibitem [{\citenamefont {Coffey}\ \emph {et~al.}(2007)\citenamefont {Coffey},
  \citenamefont {Reid}, \citenamefont {Rodovsky}, \citenamefont {Bartholomew},\
  and\ \citenamefont {Ginger}}]{Coffey2007}%
  \BibitemOpen
  \bibfield  {author} {\bibinfo {author} {\bibfnamefont {D.~C.}\ \bibnamefont
  {Coffey}}, \bibinfo {author} {\bibfnamefont {O.~G.}\ \bibnamefont {Reid}},
  \bibinfo {author} {\bibfnamefont {D.~B.}\ \bibnamefont {Rodovsky}}, \bibinfo
  {author} {\bibfnamefont {G.~P.}\ \bibnamefont {Bartholomew}}, \ and\ \bibinfo
  {author} {\bibfnamefont {D.~S.}\ \bibnamefont {Ginger}},\ }\href {\doibase
  10.1021/nl062989e} {\bibfield  {journal} {\bibinfo  {journal} {Nano Lett.}\
  }\textbf {\bibinfo {volume} {7}},\ \bibinfo {pages} {738} (\bibinfo {year}
  {2007})}\BibitemShut {NoStop}%
\bibitem [{\citenamefont {Dang}\ \emph {et~al.}(2010)\citenamefont {Dang},
  \citenamefont {Mikhailovsky},\ and\ \citenamefont {Nguyen}}]{Dang2010}%
  \BibitemOpen
  \bibfield  {author} {\bibinfo {author} {\bibfnamefont {X.~D.}\ \bibnamefont
  {Dang}}, \bibinfo {author} {\bibfnamefont {A.}~\bibnamefont {Mikhailovsky}},
  \ and\ \bibinfo {author} {\bibfnamefont {T.~Q.}\ \bibnamefont {Nguyen}},\
  }\href {\doibase 10.1063/1.3483613} {\bibfield  {journal} {\bibinfo
  {journal} {Appl. Phys. Lett.}\ }\textbf {\bibinfo {volume} {97}},\ \bibinfo
  {pages} {113303} (\bibinfo {year} {2010})}\BibitemShut {NoStop}%
\bibitem [{\citenamefont {Tennyson}\ \emph {et~al.}(2015)\citenamefont
  {Tennyson}, \citenamefont {Garrett}, \citenamefont {Frantz}, \citenamefont
  {Myers}, \citenamefont {Bekele}, \citenamefont {Sanghera}, \citenamefont
  {Munday},\ and\ \citenamefont {Leite}}]{Tennyson2015}%
  \BibitemOpen
  \bibfield  {author} {\bibinfo {author} {\bibfnamefont {E.~M.}\ \bibnamefont
  {Tennyson}}, \bibinfo {author} {\bibfnamefont {J.~L.}\ \bibnamefont
  {Garrett}}, \bibinfo {author} {\bibfnamefont {J.~A.}\ \bibnamefont {Frantz}},
  \bibinfo {author} {\bibfnamefont {J.~D.}\ \bibnamefont {Myers}}, \bibinfo
  {author} {\bibfnamefont {R.~Y.}\ \bibnamefont {Bekele}}, \bibinfo {author}
  {\bibfnamefont {J.~S.}\ \bibnamefont {Sanghera}}, \bibinfo {author}
  {\bibfnamefont {J.~N.}\ \bibnamefont {Munday}}, \ and\ \bibinfo {author}
  {\bibfnamefont {M.~S.}\ \bibnamefont {Leite}},\ }\href {\doibase
  10.1002/aenm.201501142} {\bibfield  {journal} {\bibinfo  {journal} {Adv.
  Energy Mater.}\ }\textbf {\bibinfo {volume} {5}},\ \bibinfo {pages} {1501142}
  (\bibinfo {year} {2015})}\BibitemShut {NoStop}%
\bibitem [{\citenamefont {Tennyson}\ \emph {et~al.}(2016)\citenamefont
  {Tennyson}, \citenamefont {Frantz}, \citenamefont {Howard}, \citenamefont
  {Gunnarsson}, \citenamefont {Myers}, \citenamefont {Bekele}, \citenamefont
  {Sanghera}, \citenamefont {Na},\ and\ \citenamefont {Leite}}]{Tennyson2016}%
  \BibitemOpen
  \bibfield  {author} {\bibinfo {author} {\bibfnamefont {E.~M.}\ \bibnamefont
  {Tennyson}}, \bibinfo {author} {\bibfnamefont {J.~A.}\ \bibnamefont
  {Frantz}}, \bibinfo {author} {\bibfnamefont {J.~M.}\ \bibnamefont {Howard}},
  \bibinfo {author} {\bibfnamefont {W.~B.}\ \bibnamefont {Gunnarsson}},
  \bibinfo {author} {\bibfnamefont {J.~D.}\ \bibnamefont {Myers}}, \bibinfo
  {author} {\bibfnamefont {R.~Y.}\ \bibnamefont {Bekele}}, \bibinfo {author}
  {\bibfnamefont {J.~S.}\ \bibnamefont {Sanghera}}, \bibinfo {author}
  {\bibfnamefont {S.~M.}\ \bibnamefont {Na}}, \ and\ \bibinfo {author}
  {\bibfnamefont {M.~S.}\ \bibnamefont {Leite}},\ }\href {\doibase
  10.1021/acsenergylett.6b00331} {\bibfield  {journal} {\bibinfo  {journal}
  {ACS Energy Lett.}\ }\textbf {\bibinfo {volume} {1}},\ \bibinfo {pages} {899}
  (\bibinfo {year} {2016})}\BibitemShut {NoStop}%
\bibitem [{\citenamefont {Garrett}\ \emph {et~al.}(2017)\citenamefont
  {Garrett}, \citenamefont {Tennyson}, \citenamefont {Hu}, \citenamefont
  {Huang}, \citenamefont {Munday},\ and\ \citenamefont {Leite}}]{Garrett2017}%
  \BibitemOpen
  \bibfield  {author} {\bibinfo {author} {\bibfnamefont {J.~L.}\ \bibnamefont
  {Garrett}}, \bibinfo {author} {\bibfnamefont {E.~M.}\ \bibnamefont
  {Tennyson}}, \bibinfo {author} {\bibfnamefont {M.}~\bibnamefont {Hu}},
  \bibinfo {author} {\bibfnamefont {J.}~\bibnamefont {Huang}}, \bibinfo
  {author} {\bibfnamefont {J.~N.}\ \bibnamefont {Munday}}, \ and\ \bibinfo
  {author} {\bibfnamefont {M.~S.}\ \bibnamefont {Leite}},\ }\href {\doibase
  10.1021/acs.nanolett.7b00289} {\bibfield  {journal} {\bibinfo  {journal}
  {Nano Lett.}\ }\textbf {\bibinfo {volume} {17}},\ \bibinfo {pages} {2554}
  (\bibinfo {year} {2017})}\BibitemShut {NoStop}%
\bibitem [{\citenamefont {Cho}(2011)}]{Cho2011}%
  \BibitemOpen
  \bibfield  {author} {\bibinfo {author} {\bibfnamefont {Y.}~\bibnamefont
  {Cho}},\ }\href {\doibase 10.1016/S1076-5670(03)80096-4} {\bibfield
  {journal} {\bibinfo  {journal} {J. Mater. Res.}\ }\textbf {\bibinfo {volume}
  {26}},\ \bibinfo {pages} {2007} (\bibinfo {year} {2011})}\BibitemShut
  {NoStop}%
\bibitem [{\citenamefont {Williams}\ \emph {et~al.}(1989)\citenamefont
  {Williams}, \citenamefont {Slinkman}, \citenamefont {Hough},\ and\
  \citenamefont {Wickramasinghe}}]{Williams1989}%
  \BibitemOpen
  \bibfield  {author} {\bibinfo {author} {\bibfnamefont {C.~C.}\ \bibnamefont
  {Williams}}, \bibinfo {author} {\bibfnamefont {J.}~\bibnamefont {Slinkman}},
  \bibinfo {author} {\bibfnamefont {W.~P.}\ \bibnamefont {Hough}}, \ and\
  \bibinfo {author} {\bibfnamefont {H.~K.}\ \bibnamefont {Wickramasinghe}},\
  }\href {\doibase 10.1063/1.102312} {\bibfield  {journal} {\bibinfo  {journal}
  {Appl. Phys. Lett.}\ }\textbf {\bibinfo {volume} {55}},\ \bibinfo {pages}
  {1662} (\bibinfo {year} {1989})}\BibitemShut {NoStop}%
\bibitem [{\citenamefont {Cui}\ \emph {et~al.}(2016)\citenamefont {Cui},
  \citenamefont {Ma},\ and\ \citenamefont {Shen}}]{Cui2016}%
  \BibitemOpen
  \bibfield  {author} {\bibinfo {author} {\bibfnamefont {Y.~T.}\ \bibnamefont
  {Cui}}, \bibinfo {author} {\bibfnamefont {E.~Y.}\ \bibnamefont {Ma}}, \ and\
  \bibinfo {author} {\bibfnamefont {Z.~X.}\ \bibnamefont {Shen}},\ }\href
  {\doibase 10.1063/1.4954156} {\bibfield  {journal} {\bibinfo  {journal} {Rev.
  Sci. Instrum.}\ }\textbf {\bibinfo {volume} {87}} (\bibinfo {year} {2016}),\
  10.1063/1.4954156}\BibitemShut {NoStop}%
\bibitem [{\citenamefont {Sviridov}\ and\ \citenamefont
  {Kozlovsky}(2016)}]{sviridov2016}%
  \BibitemOpen
  \bibfield  {author} {\bibinfo {author} {\bibfnamefont {D.~E.}\ \bibnamefont
  {Sviridov}}\ and\ \bibinfo {author} {\bibfnamefont {V.~I.}\ \bibnamefont
  {Kozlovsky}},\ }\href {\doibase 10.1116/1.4964713} {\bibfield  {journal}
  {\bibinfo  {journal} {J. Vac. Sci. Technol. B, Nanotechnol. Microelectron.
  Mater. Process. Meas. Phenom.}\ }\textbf {\bibinfo {volume} {34}},\ \bibinfo
  {pages} {061801} (\bibinfo {year} {2016})}\BibitemShut {NoStop}%
\bibitem [{\citenamefont {Bussmann}\ and\ \citenamefont
  {Williams}(2004)}]{Bussmann2004}%
  \BibitemOpen
  \bibfield  {author} {\bibinfo {author} {\bibfnamefont {E.}~\bibnamefont
  {Bussmann}}\ and\ \bibinfo {author} {\bibfnamefont {C.~C.}\ \bibnamefont
  {Williams}},\ }\href {\doibase 10.1063/1.1641161} {\bibfield  {journal}
  {\bibinfo  {journal} {Rev. Sci. Instrum.}\ }\textbf {\bibinfo {volume}
  {75}},\ \bibinfo {pages} {422} (\bibinfo {year} {2004})}\BibitemShut
  {NoStop}%
\bibitem [{\citenamefont {Xiong}\ \emph {et~al.}(2010)\citenamefont {Xiong},
  \citenamefont {Lu}, \citenamefont {Zhou}, \citenamefont {Jiang},
  \citenamefont {Wang}, \citenamefont {Qiu}, \citenamefont {Dong},\ and\
  \citenamefont {Yang}}]{Xiong2010}%
  \BibitemOpen
  \bibfield  {author} {\bibinfo {author} {\bibfnamefont {K.}~\bibnamefont
  {Xiong}}, \bibinfo {author} {\bibfnamefont {S.}~\bibnamefont {Lu}}, \bibinfo
  {author} {\bibfnamefont {T.}~\bibnamefont {Zhou}}, \bibinfo {author}
  {\bibfnamefont {D.}~\bibnamefont {Jiang}}, \bibinfo {author} {\bibfnamefont
  {R.}~\bibnamefont {Wang}}, \bibinfo {author} {\bibfnamefont {K.}~\bibnamefont
  {Qiu}}, \bibinfo {author} {\bibfnamefont {J.}~\bibnamefont {Dong}}, \ and\
  \bibinfo {author} {\bibfnamefont {H.}~\bibnamefont {Yang}},\ }\href {\doibase
  10.1016/j.solener.2010.05.013} {\bibfield  {journal} {\bibinfo  {journal}
  {Sol. Energy}\ }\textbf {\bibinfo {volume} {84}},\ \bibinfo {pages} {1888}
  (\bibinfo {year} {2010})}\BibitemShut {NoStop}%
\bibitem [{\citenamefont {Keavney}\ \emph {et~al.}(1990)\citenamefont
  {Keavney}, \citenamefont {Haven},\ and\ \citenamefont {Vernon}}]{111606}%
  \BibitemOpen
  \bibfield  {author} {\bibinfo {author} {\bibfnamefont {C.~J.}\ \bibnamefont
  {Keavney}}, \bibinfo {author} {\bibfnamefont {V.~E.}\ \bibnamefont {Haven}},
  \ and\ \bibinfo {author} {\bibfnamefont {S.~M.}\ \bibnamefont {Vernon}},\
  }in\ \href {\doibase 10.1109/PVSC.1990.111606} {\emph {\bibinfo {booktitle}
  {IEEE Conf. Photovolt. Spec.}}}\ (\bibinfo {year} {1990})\ pp.\ \bibinfo
  {pages} {141--144}\BibitemShut {NoStop}%
\bibitem [{\citenamefont {Liang}\ \emph {et~al.}(2015)\citenamefont {Liang},
  \citenamefont {Han}, \citenamefont {Fan},\ and\ \citenamefont
  {Xing}}]{Liang2015}%
  \BibitemOpen
  \bibfield  {author} {\bibinfo {author} {\bibfnamefont {P.}~\bibnamefont
  {Liang}}, \bibinfo {author} {\bibfnamefont {P.-D.}\ \bibnamefont {Han}},
  \bibinfo {author} {\bibfnamefont {Y.-J.}\ \bibnamefont {Fan}}, \ and\
  \bibinfo {author} {\bibfnamefont {Y.-P.}\ \bibnamefont {Xing}},\ }\href
  {\doibase 10.1088/1674-1056/24/3/038801} {\bibfield  {journal} {\bibinfo
  {journal} {Chinese Phys. B}\ }\textbf {\bibinfo {volume} {24}},\ \bibinfo
  {pages} {038801} (\bibinfo {year} {2015})}\BibitemShut {NoStop}%
\bibitem [{\citenamefont {Zhao}\ \emph {et~al.}(1996)\citenamefont {Zhao},
  \citenamefont {Wang}, \citenamefont {Altermatt}, \citenamefont {Wenham},\
  and\ \citenamefont {Green}}]{Zhao1996}%
  \BibitemOpen
  \bibfield  {author} {\bibinfo {author} {\bibfnamefont {J.}~\bibnamefont
  {Zhao}}, \bibinfo {author} {\bibfnamefont {A.}~\bibnamefont {Wang}}, \bibinfo
  {author} {\bibfnamefont {P.~P.}\ \bibnamefont {Altermatt}}, \bibinfo {author}
  {\bibfnamefont {S.~R.}\ \bibnamefont {Wenham}}, \ and\ \bibinfo {author}
  {\bibfnamefont {M.~A.}\ \bibnamefont {Green}},\ }\href {\doibase
  10.1016/0927-0248(95)00117-4} {\bibfield  {journal} {\bibinfo  {journal}
  {Sol. Energy Mater. Sol. Cells}\ }\textbf {\bibinfo {volume} {41-42}},\
  \bibinfo {pages} {87} (\bibinfo {year} {1996})}\BibitemShut {NoStop}%
\bibitem [{\citenamefont {Holman}\ \emph {et~al.}(2013)\citenamefont {Holman},
  \citenamefont {Descoeudres}, \citenamefont {{De Wolf}},\ and\ \citenamefont
  {Ballif}}]{Holman2013}%
  \BibitemOpen
  \bibfield  {author} {\bibinfo {author} {\bibfnamefont {Z.}~\bibnamefont
  {Holman}}, \bibinfo {author} {\bibfnamefont {a.}~\bibnamefont {Descoeudres}},
  \bibinfo {author} {\bibfnamefont {S.}~\bibnamefont {{De Wolf}}}, \ and\
  \bibinfo {author} {\bibfnamefont {C.}~\bibnamefont {Ballif}},\ }\href
  {\doibase 10.1109/JPHOTOV.2013.2276484} {\bibfield  {journal} {\bibinfo
  {journal} {IEEE J. Photovoltaics}\ }\textbf {\bibinfo {volume} {3}},\
  \bibinfo {pages} {1243} (\bibinfo {year} {2013})}\BibitemShut {NoStop}%
\bibitem [{\citenamefont {Kiliani}\ \emph {et~al.}(2011)\citenamefont
  {Kiliani}, \citenamefont {Micard}, \citenamefont {Steuer}, \citenamefont
  {Raabe}, \citenamefont {Herguth},\ and\ \citenamefont {Hahn}}]{Kiliani2011}%
  \BibitemOpen
  \bibfield  {author} {\bibinfo {author} {\bibfnamefont {D.}~\bibnamefont
  {Kiliani}}, \bibinfo {author} {\bibfnamefont {G.}~\bibnamefont {Micard}},
  \bibinfo {author} {\bibfnamefont {B.}~\bibnamefont {Steuer}}, \bibinfo
  {author} {\bibfnamefont {B.}~\bibnamefont {Raabe}}, \bibinfo {author}
  {\bibfnamefont {A.}~\bibnamefont {Herguth}}, \ and\ \bibinfo {author}
  {\bibfnamefont {G.}~\bibnamefont {Hahn}},\ }\href {\doibase
  10.1063/1.3630031} {\bibfield  {journal} {\bibinfo  {journal} {J. Appl.
  Phys.}\ }\textbf {\bibinfo {volume} {110}},\ \bibinfo {pages} {054508}
  (\bibinfo {year} {2011})}\BibitemShut {NoStop}%
\end{thebibliography}%


\appendix

\section{Intensity Normalization \label{intensity}}

\begin{figure}
\includegraphics[width = 7cm]{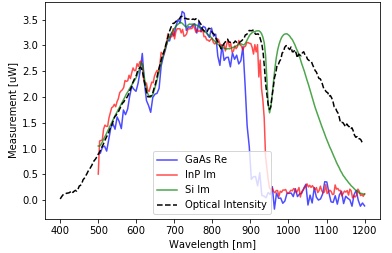}
\caption{
Thermopile measurement of optical intensity overlaid with raw MIM spectra for Si, GaAs and InP.
}
\label{figS1}
\end{figure}

The monochrometer output intensity varies significantly over wavelength, and the resulting MIM response varies with it, as shown in Fig. \ref{figS1}. Thus, the MIM signal is normalized by this intensity, as shown in Fig. 2 of the main text. This normalization assumes that the MIM signal varies linearly with applied optical intensity, which is only a good assumption because the optical intensities used are so small, resulting in small changes in conductivity. Larger changes in conductivity, such as those arising from a more powerful optical source, would require a more sophisticated normalization taking into account the nonlinear dependence of MIM signal on conductivity.\\

\section{Topography Effects in Highly Textured Silicon\label{texture}}

\begin{figure}
\includegraphics[width = 8cm]{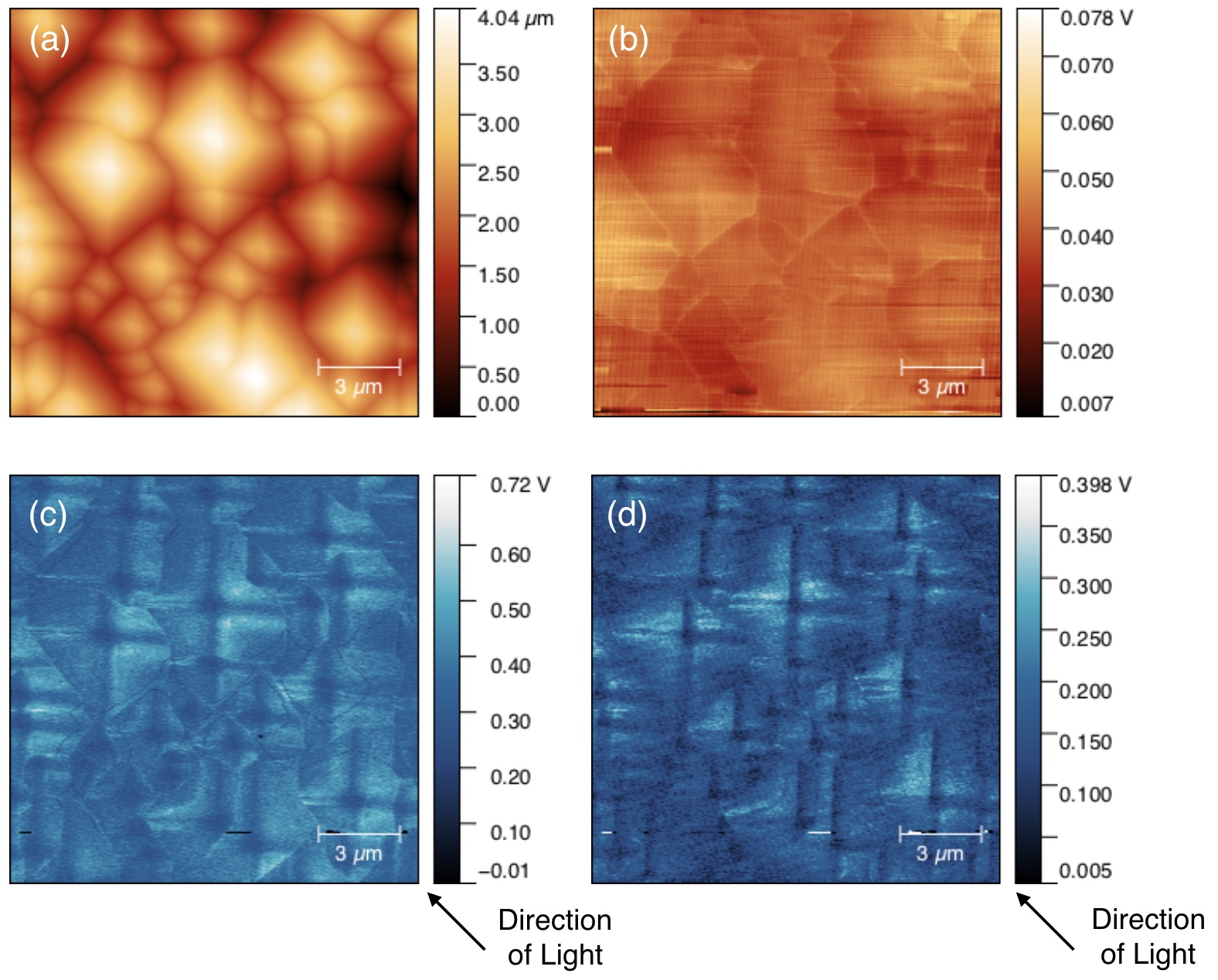}
\caption{
Topography (a), MIM-Re (b), optically modulated MIM-Im (c), and optically modulated MIM-Re (d) of highly textured Si. 
}
\label{figS2}
\end{figure}

In order to evaluate the effects of topography on optically modulated MIM measurement we measured textured, lightly doped, mono-crystalline silicon. Since we expect the photoconductivity of the sample to be relatively uniform, the non-uniform signal measured should be a result of only topographical artifacts. As shown in Fig. \ref{figS2}, these artifacts appear to be very significant. The artifacts in MIM-Re are easy to explain due to the geometry of the tip-sample interface. The artifacts in optically modulated MIM are less easily explained, however, and could be a result of inhomogenous illumination. These artifacts appear to be a general issue for top-illuminated electrical scanning probe measurements of highly textured samples.

\end{document}